\documentclass[
  a4paper,
  fontsize=11pt,
  ]{article}

\pdfoutput=1

\usepackage{ILD}
\usepackage{graphicx}

\usepackage[utf8]{inputenx}
\usepackage[T1]{fontenc}
\usepackage[british]{babel}
\usepackage{csquotes}

\usepackage{subcaption}
\usepackage{import}
\usepackage{xspace}
\usepackage{graphicx}
\usepackage[shortcuts]{extdash}
\usepackage{siunitx}
\DeclareSIUnit \lightspeed {\text{{c}}}
\usepackage{xcolor}
\definecolor{linkblue}{HTML}{264772}
\usepackage{hyperref}
\hypersetup{
    colorlinks=true,
    linktocpage=true,
    linkcolor=linkblue,
    citecolor=linkblue,
    urlcolor=linkblue
}

\usepackage{wrapfig}

\graphicspath{ {../pictures/} }
\DeclareGraphicsExtensions{.pdf,.png,.jpg}

\newcommand{\dEdx}{\ensuremath{\mathrm{d}E/\mathrm{d}x}\xspace}
\newcommand{\dedx}{\dEdx}

\begin{document}

\hyphenation{
  am-pli-fi-ca-tion
  col-lab-o-ra-tion
  per-for-mance
  sat-u-rat-ed
  se-lect-ed
  spec-i-fied
}

\title{Charged Hadron Identification with dE/dx and Time-of-Flight at Future Higgs Factories}
\addauthor{Ulrich Einhaus}{\institute{1}\institute{2}}
\addinstitute{1}{Deutsches Elektronen-Synchrotron DESY, Notkestr. 85, 22607 Hamburg, Germany}
\addinstitute{2}{Universität Hamburg, Department of Physics, Jungiusstraße 9, 20355 Hamburg, Germany}


\abstract{
The design of detector concepts has been driven for a long time by requirements on transverse momentum, impact parameter and jet energy resolutions, as well as hermeticity. Only rather recently it has been realised that the ability to idenfity different types of charged hadrons, in particular kaons and protons, could have important applications at Higgs factories like the International Linear Collider (ILC), ranging from improvements in tracking, vertexing and flavour tagging to measurements requiring strangeness-tagging. While detector concepts with gaseous tracking, like a time projection chamber (TPC), can exploit the specific energy loss, all-silicon-based detectors have to rely on fast timing layers in front of or in the first layers of their electromagnetic calorimeters (ECals). This work will review the different options for realising particle identification (PID) for pions, kaons and protons, introduce recently developed reconstruction algorithms and present full detector simulation prospects for physics applications using the example of the International Large Detector (ILD) concept.
}

\titlecomment{This work was carried out in the framework of the ILD detector concept group.}

\ildproc{phys}{2021}{015}

\titlepage


\newpage
\section{Introduction}
\begin{figure}[!hbt]
  \centering
    \includegraphics[width=.85\textwidth,keepaspectratio=true]{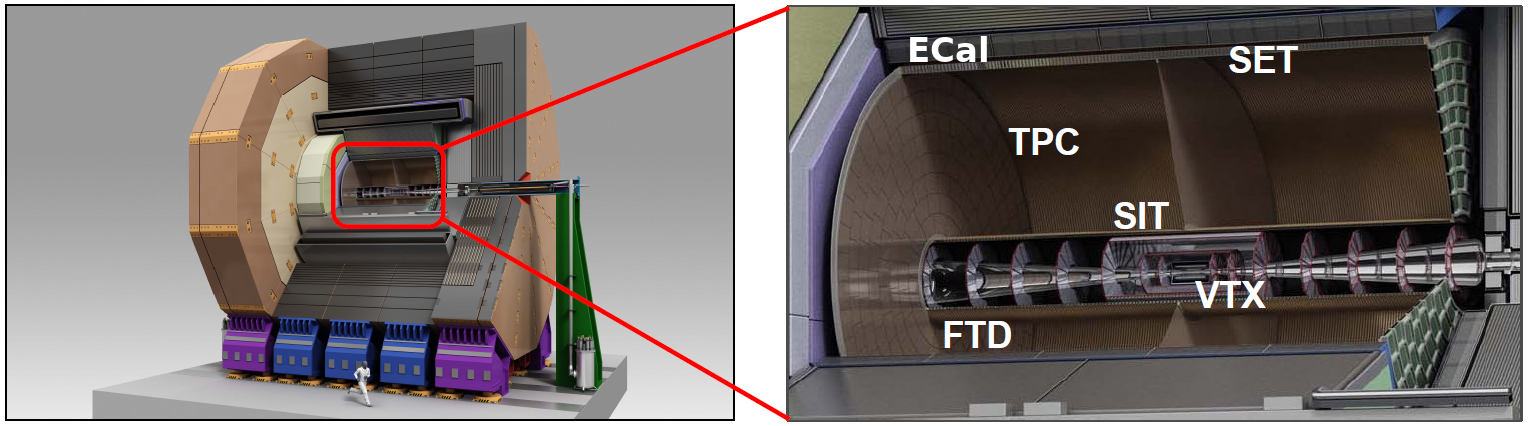}
    \hspace{.1cm}
    \includegraphics[width=.13\textwidth,keepaspectratio=true]{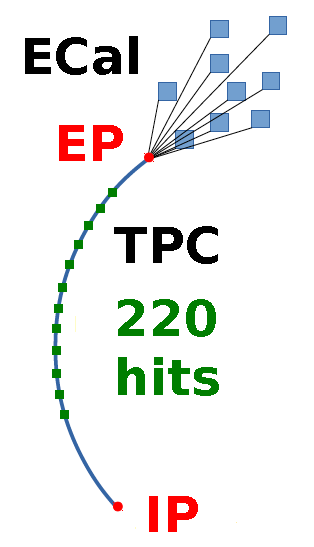}
  \centering
    \caption{Schematic view of ILD, from \cite{ILD_IDR}, and scheme of hits in the TPC and the Ecal used for \dedx and TOF measurements, repectively.}
  \label{fig:ILD}
\end{figure}

The ILC \cite{ILC_TDR_Summary} is a proposed \SI{250}{} - \SI{500}{GeV} $e^+e^-$-collider and ILD \cite{ILD_IDR} is one of its detector concepts, shown in \autoref{fig:ILD}.
It is a multi-purpose detector with a silicon vertex tracker (VTX), a TPC with a silicon envelope (SIT, SET) as central tracking system and a highly granular calorimeter system inside a \SI{3.5}{T} solenoid and a muon system outside of it.
With its forward tracker (FTD) and forward calorimeter system it achieves a high degree of hermeticity.
ILD is designed for particle flow \cite{Pandora} and has an asymptotic momentum resolution of \SI{2e-5}{GeV^{-1}} and a jet energy resolution of better than \SI{3.5}{\percent} above \SI{100}{GeV}.
This work concentrates on the PID capabilities of ILD via measurement of the time-of-flight (TOF) in the ECal in full-detector simulation \cite{ILD_IDR}, while PID via measurement of the specific energy loss \dedx in the TPC has been discussed separately \cite{Einhaus_ProcEPS_2021}.
This work makes use of a large MC production in 2018 \cite{Production_2018}, which generated, simulated and reconstructed about \SI{500}{fb^{-1}} of ILC integrated luminosity.
This includes the entire Standard Model processes and single particles for calibration and detailed studies.

\section{Time-of-Flight Measurement TOF}

\begin{figure}
\centering
\begin{minipage}{.5\textwidth}
  \centering
    \includegraphics[width=.96\textwidth,keepaspectratio=true]{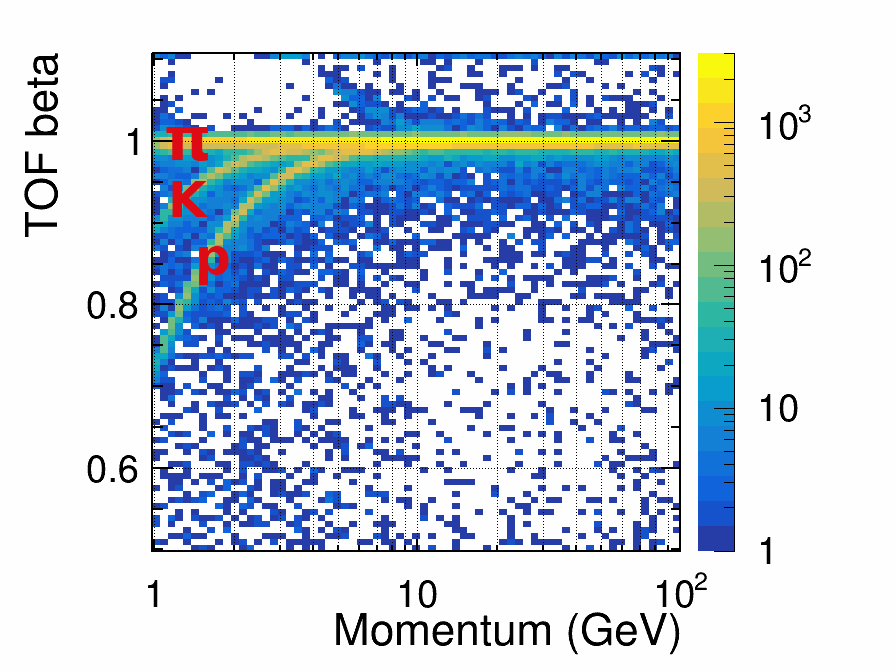}
  \captionof{figure}{TOF $\beta$ curves of particles in single particle events.}
  \label{fig:TOF_beta}
\end{minipage}
\hspace{.2cm}
\begin{minipage}{.46\textwidth}
  \centering
  \includegraphics[width=.96\textwidth,keepaspectratio=true]{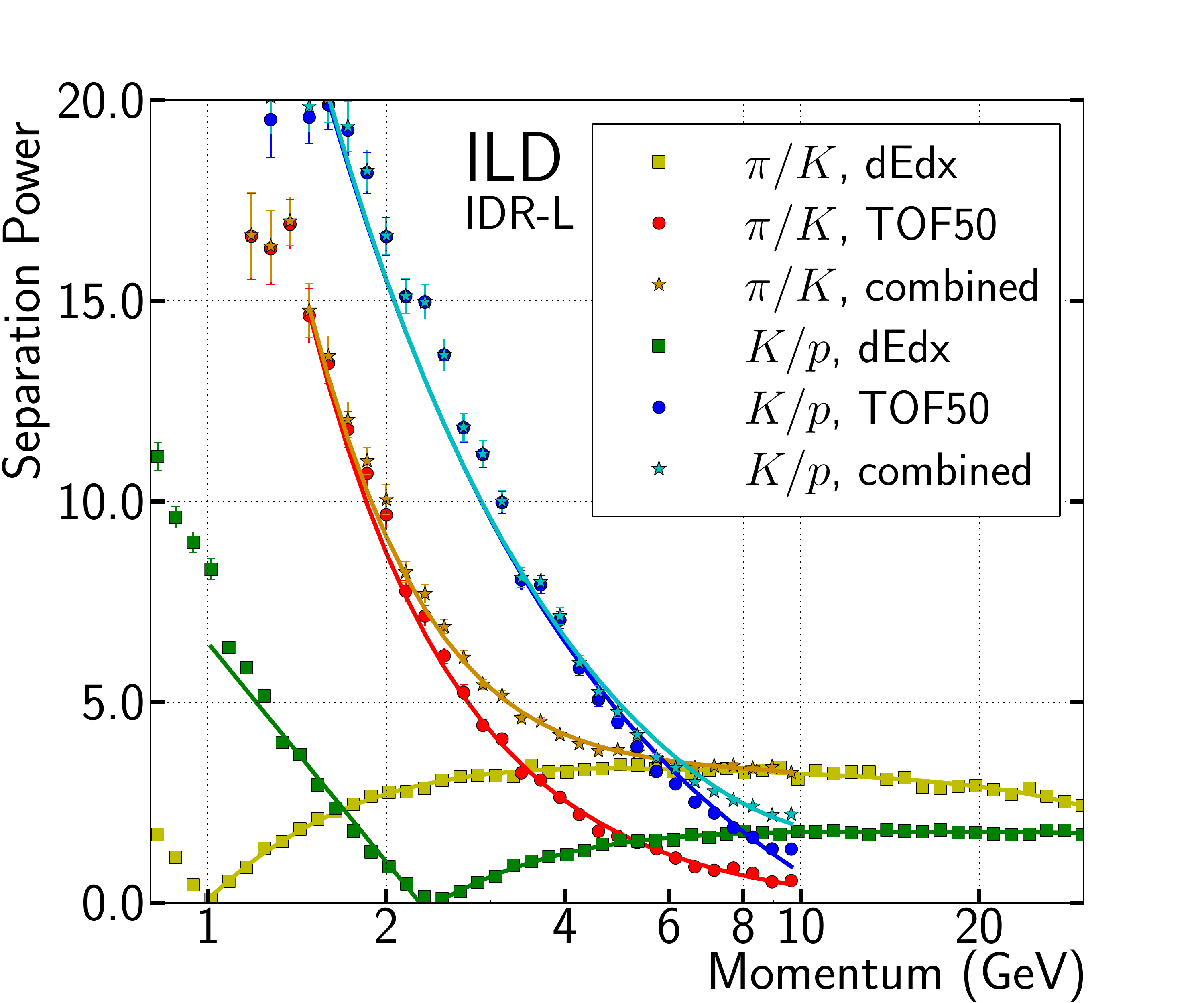}
  \captionof{figure}{Combined TOF and \dedx separation power for pion/kaon and kaon/proton separation. For details on \dedx see \cite{Einhaus_ProcEPS_2021}. The curves are only to guide the eye.}
  \label{fig:SP_comb}
\end{minipage}
\end{figure}

For the TOF measurement, a timing resolution of \SI{50}{ps} per channel has been assumed to be achievable for the ECal and was implemented in the simulation.
The TOF estimator for an incident particle is calculated using the first 10 layers of the ECal.
The timing values of the one active channel in each layer which is closest to the extrapolated track are projected back to the entry point (EP) into the ECal assuming propagation with the speed of light and then averaged.
Together with the track length, the absolute velocity of the particle $\beta$ in units of $c$ is calculated.
This $\beta$ is shown in \autoref{fig:TOF_beta}, with bands of pions, kaons and protons which are well separable up to \SI{3}{GeV} for $\pi/K$ and \SI{6}{GeV} for $K/p$.
The separation power $S$ is the relative distance between the bands, defined as
$S = |\mu_1-\mu_2|/\sqrt{\frac{\sigma_1^2+\sigma_2^2}{2}}$
with $\mu_i$ and $\sigma_i$ being the mean and width of the band of particle $i$, respectively.
This separation power can be calculated and combined with the one from \dedx \cite{Einhaus_ProcEPS_2021} in quadrature, which is shown in \autoref{fig:SP_comb}.

\begin{figure}
\centering
\begin{minipage}{.48\textwidth}
  \centering
    \includegraphics[width=.96\textwidth,keepaspectratio=true]{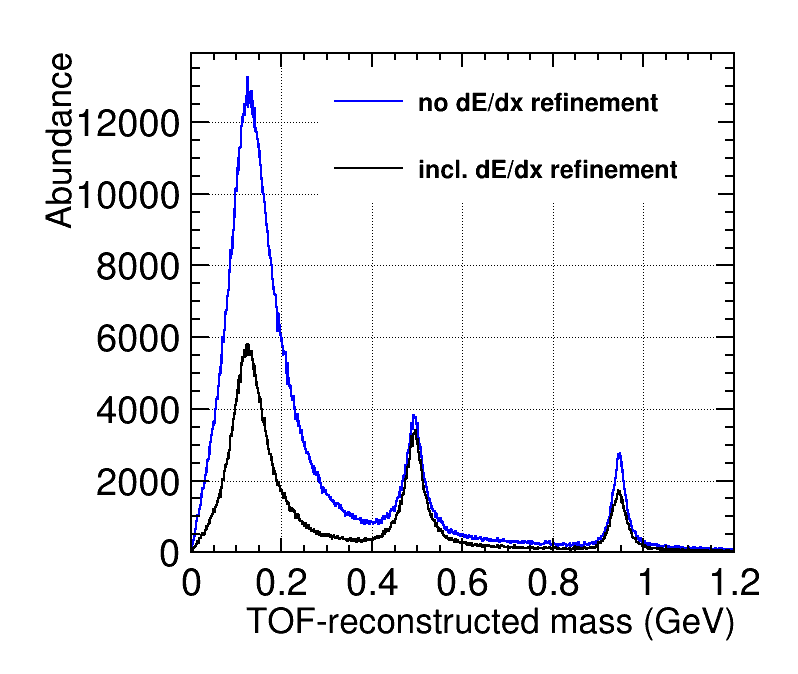}
  \captionof{figure}{TOF-reconstructed masses in full physics events. Pions, kaons and protons are well separable, and \dedx refinement can be used to improve kaon selection.}
  \label{fig:Reco_Mass}
\end{minipage}
\hspace{.2cm}
\begin{minipage}{.48\textwidth}
  \centering
  \includegraphics[width=.96\textwidth,keepaspectratio=true]{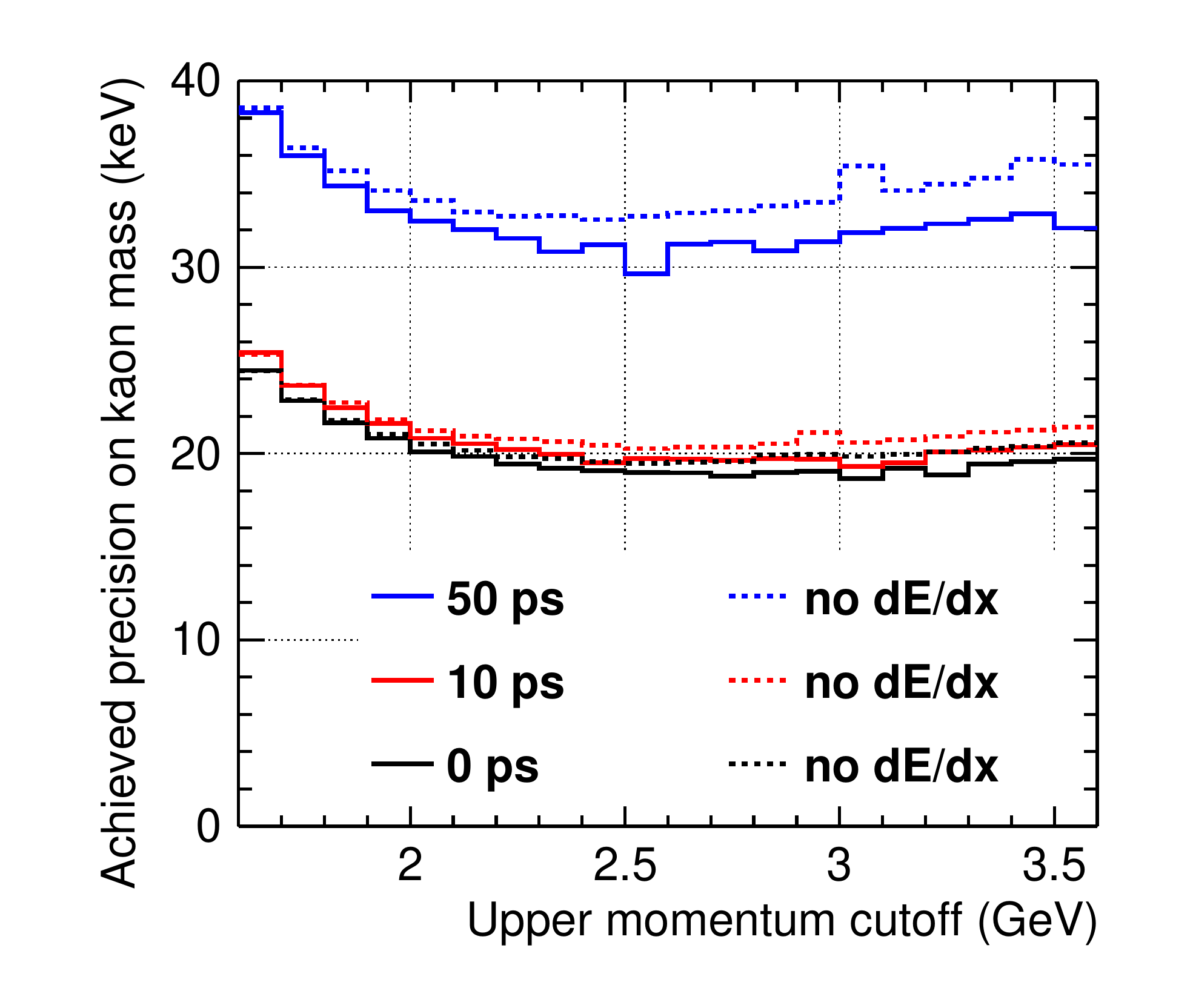}
  \captionof{figure}{Statistical precision on the kaon mass depending on the simulated TOF timing resolution for \SI{200}{fb^{-1}} of ILC at \SI{500}{GeV}.}
  \label{fig:KaonMass}
\end{minipage}
\end{figure}

A possible application of TOF is the measurement of the charged kaon mass \cite{Einhaus21}.
This quantity is known to a precision of $m_K = (493,677 \pm 13)$ \SI{}{keV}, but the two most precise measurements disagree with each other by \SI{60}{keV} with individual error bars of about \SI{10}{keV} \cite{PDG2020}.
In ILD physics events, the mass $m$ of a particle can be calculated using the reconstructed TOF $\beta$ and the measured momentum $p$ via $m=p\sqrt{1/\beta^2-1}$, which is shown in \autoref{fig:Reco_Mass}.
Here, a timing resolution of \SI{50}{ps} and only particles with $p < \SI{3}{GeV}$ (upper momentum cut) are used.
The kaon peak is well identifiable and is fitted.
To reduce background, a particle's \dedx value must be consistent with the kaon hypothesis within 2.5$\sigma$ (black histogram).
In \autoref{fig:KaonMass} the resulting statistical uncertainty on the fitted kaon mass is shown for an integrated ILC luminosity of \SI{200}{fb^{-1}} at \SI{500}{GeV}.
The precision can be optimised with respect to the chosen upper momentum cut and depends on the simulated timing resolution.
With a realistic timing resolution of \SI{50}{ps}, a statistical uncertainty of \SI{30}{keV} or \SI{6e-5}{} can be achieved, which motivates studying systematic effects in the future.

\section{TOF Ongoing Development}

\begin{figure}
\centering
\begin{minipage}{.48\textwidth}
  \centering
    \includegraphics[width=.96\textwidth,keepaspectratio=true]{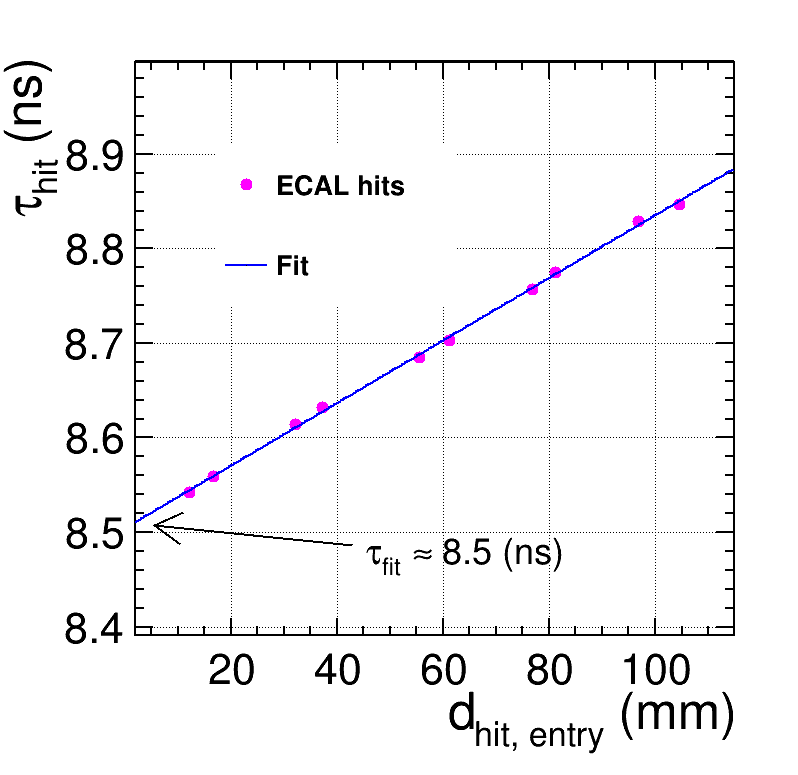}
  \captionof{figure}{Example of the TOF fit method. Instead of assuming a propagation velocity c in the ECal, this velocity is fitted and gives a TOF estimator at the EP. From \cite{Timing_2021}.}
  \label{fig:TOF_FitMethod}
\end{minipage}
\hspace{.2cm}
\begin{minipage}{.48\textwidth}
  \centering
  \includegraphics[width=.96\textwidth,keepaspectratio=true]{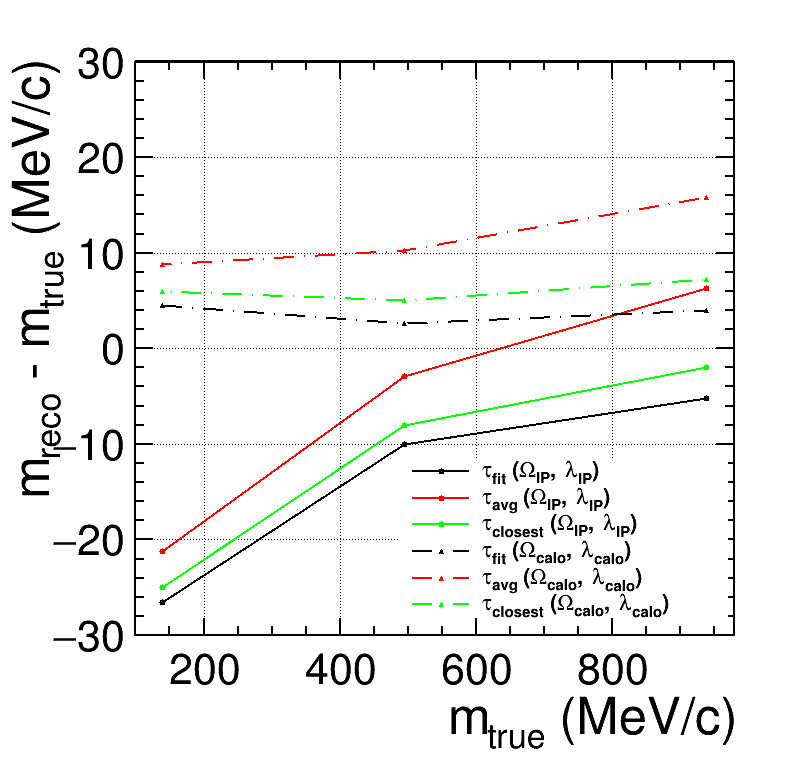}
  \captionof{figure}{Bias of the reconstructed masses of pions, kaons and protons, depending on the TOF algorithm $\tau$ and the reference point for the momentum reconstruction, IP calorimeter EP. From \cite{Timing_2021}.}
  \label{fig:RecoMassImprove}
\end{minipage}
\end{figure}

The largest systematic uncertainty on a mass measurement comes from the TOF algorithm and is in the order of \SI{10}{MeV}.
Studies are ongoing to improve the algorithm and reduce this bias \cite{Timing_2021}.
The developments will be implemented in a future MC production and include the following aspects.
Instead of projecting the measured time of the individual ECal hits to the EP using a propagation velocity c, this velocity can be fitted, compare the example in  \autoref{fig:TOF_FitMethod}.
For extended showers in the ECal, the cluster timing can be calibrated based on the number of cluster hits, which is an effect at the level of $10^{-3}$.
Since a particle loses energy in the tracker, the reconstructed momentum at the calorimeter EP can be used instead of the one at the IP. This changes the sign of the bias, but reduces its size, as displayed in \autoref{fig:RecoMassImprove}.
Here, also 3 TOF algorithms are compared: the default one $\tau_\mathrm{avg}$, the fit method $\tau_\mathrm{fit}$, and using only one ECal hit per particle, namely the one closest to the EP, $\tau_\mathrm{closest}$.
The combination with the smallest bias across the three studied hadron masses is using the momentum at the EP and $\tau_\mathrm{fit}$ (black dashed line).

\section{Conclusions}

With novel timing technologies, TOF provides a new opportunity for PID at high energy colliders.
Measurements of TOF and \dedx provide sensitivity for $\pi/K$ and $K/p$ separation in complementary momentum ranges, in particular TOF excels in the `blind spots' of \dedx.
With TOF, a measurement of the charge kaon mass at the level of better than $10^{-4}$ is statistically achievable, provided the systematic uncertainties from the TOF measurement and the momentum scale can be kept at the same level.
This measurement would help to solve the long-standing disagreement of the kaon mass.
Studies are ongoing to improve the systematic precision of the TOF estimator algorithm by making it more realistic.
Recent developments improved the bias on the reconstructed hadron mass from several \SI{10}{MeV} to several \SI{}{MeV}.


\bibliographystyle{JHEP}
\bibliography{References_reduced}
\end{document}